\documentclass[submission, Phys]{SciPost}

\hypersetup{
    colorlinks,
    linkcolor={red!50!black},
    citecolor={blue!50!black},
    urlcolor={blue!80!black}
}

\usepackage[bitstream-charter]{mathdesign}
\usepackage{physics}
\usepackage{xcolor}
\DeclareSymbolFont{usualmathcal}{OMS}{cmsy}{m}{n}

\DeclareSymbolFontAlphabet{\mathcal}{usualmathcal}

\begin{document}

\begin{center}{\Large \textbf{
Dynamics of the order parameter statistics in the long range Ising model\\
}}\end{center}

\begin{center}
Nishan Ranabhat\textsuperscript{1,2$\star$},
Mario Collura\textsuperscript{1}
\end{center}

\begin{center}
{\bf 1} SISSA - International School for Advanced Studies, via Bonomea 265, 34136 Trieste, Italy
\\
{\bf 2} The Abdus Salam International Centre for Theoretical Physics,
Strada Costiera 11, 34151 Trieste, Italy
\\
${}^\star$ {\small \sf nranabha@sissa.it}
\end{center}

\begin{center}
\today
\end{center}


\section*{Abstract}
{\bf
We study the relaxation of the local ferromagnetic order in the transverse field quantum Ising chain with power-law decaying interactions $\sim 1/r^\alpha$. We prepare the system in the GHZ state and study the time evolution of the probability distribution function (PDF) of the  order parameter within a block of $\ell$ when quenching the transverse field. The model is known to support long range order at finite temperature for $\alpha\leq 2.0$. In this regime, quasi-localized topological magnetic defects are expected to strongly affect the equilibration of the full probability distribution. 
We highlight different dynamical regimes where gaussification mechanism may be slowed down by confinement and eventually breaks. 
We further study the PDF dynamics induced by changing the effective dimensionality of the system; we mimic this by quenching the range of the interactions. As a matter of fact, the behavior of the system crucially depends on the value of $\alpha$ governing the unitary evolution.
}

\vspace{10pt}
\noindent\rule{\textwidth}{1pt}
\tableofcontents\thispagestyle{fancy}
\noindent\rule{\textwidth}{1pt}
\vspace{10pt}

\section{Introduction}
\label{sec:intro}
Advances in the field of synthetic quantum matter in laboratory has allowed us to gain deeper insights into the static and dynamic properties of isolated many body quantum systems
\cite{hundred,Quench_1D_Atm_Isi,UC_optical_lattice,UC_out_eq,NC,DG,Ramsey_intferro,quant_PT_expt}. One of the primary objective of these experiments is to calculate the full quantum mechanical probability distribution of certain observables as it holds the full information about the quantum fluctuation in the system. In parallel, several analytical and numerical studies have been carried out to study the full probability distribution of the system's order parameter both in equilibrium and out of equilibrium regime \cite{FCS_ising,order_melt,relax,FCS_XXZ,dyn_con,quant_noise,fcs_crit_ising,fcs_XY,fcs_Hal_sha}. In this work we study the dynamics of order parameter statistics, encoded in its probability distribution, of the long range Ising model following a global quantum quench \cite{quench}. Similar studies have been carried out for systems with short range interaction \cite{FCS_ising,order_melt,relax,dyn_con}, where the main objective is to see if and how the initial order in the system melts after quantum quench. long range Ising model is more interesting as it exhibits short-range Ising and mean-field universality class \cite{Ramsey_intferro} connected by an intermediate region where the critical exponents vary monotonically \cite{fidelity}. Quenches in the mean-field regime reveals information on the dynamics of systems with dimension higher than one. 

We perform two different kind of quenches; the first kind is along the transverse field at constant value of interaction range. Here we initialize our system deep in the ferromagnetic region, which is characterized by double peak probability distribution function (PDF) of our local order parameter. Locality is a crucial concept to ensure the relaxation of the order parameter at late time \cite{Essler_2016}. The system is then suddenly quenched along the transverse field and evolved with the new Hamiltonian. We are interested in how the initial ferromagnetic order melts after a quantum quench and specially if the PDF attains a Gaussian shape around zero in the maximum time limit and system size that we can effectively simulate. The second kind of quench is performed along the direction of the interaction range, keeping constant transverse field. We perform quenches in both directions: initializing the system as fully connected and quenching to a short range Hamiltonian, and vice versa. We investigate how the system retains memory of the initial ferromagnetic order after the quench, and whether quenches performed in opposite directions are qualitatively equivalent.

Specifically, the manuscript is organized as follows:

\begin{itemize}
  \item In Sec. \ref{sec:model} we introduce the model, the parameters involved, and the equilibrium phase diagram. We define the local order parameter and introduce the full counting statistics. We define the cumulant generating function and its relation to the probability distribution function (PDF) and outline the steps to compute each of these in a spin model. We define the general quench protocol and the algorithms used.
  
  \item Sec.~\ref{sec:res} is devoted to the results. We present the results for the dynamics of order parameter statistics for quenches along the transverse field, at different values of the interaction range. We show that there is a region in the parameter space where the the system never shows Gaussification, independently of the simulation time and the system size. On the contrary, outside this region the Gaussification of the PDF strongly depends on the simulation parameters. We also present the results for the dynamics of the order parameter statistics for quenches along the interaction range at different values of transverse field. Here we show that the dynamics is strongly dependent on the post quench parameters and that the quenches performed in the opposite direction yields qualitatively different behaviour.
  
  \item In Sec. \ref{sec:conclude} we draw our conclusion, also mentioning further possible lines of investigation and eventually connecting this study to finite temperature analysis.
\end{itemize}

\section{Model and methods}\label{sec:model}

\subsection{Hamiltonian of long range transverse field Ising model}
We study the long range transverse field Ising model. A spin model like this can be experimentally realized with systems of trapped ions \cite{Ramsey_intferro,trapped_ion1,trapped_ion2} where the interaction is mediated by collective vibrations. The Hamiltonian is

\begin{equation}
\label{eq:mdl1}
    H(\alpha,h/J) = -\frac{1}{\mathcal{K}(\alpha)}\sum_{i<j}^N \frac{J}{\vert i-j\vert^\alpha} \hat{s}^x_i \hat{s}^x_j - h \sum_{i=1}^N \hat{s}^z_i
\end{equation}

where $\hat{s}^a_i, a = x,y,z$ are the spin $1/2$ matrices on the $i^{th}$ lattice site. The spin-spin interaction is long ranged and is decaying as the inverse power of the distance between two spins. This interaction is tuned by the interaction range $\alpha$. The Kac normalization is defined as

\begin{equation} \label{eq:mdl2}
    \mathcal{K}(\alpha) = \frac{1}{N-1}\sum_{i<j}^N \frac{J}{\vert i-j\vert^\alpha} = \frac{1}{N-1}\sum_{n=1}^N \frac{N-n}{n^\alpha}
\end{equation}
and ensures the intensive scaling of the energy density for $\alpha < 1$, which otherwise blows up since the spin-spin interaction series becomes hyper-harmonic in the thermodynamic limit.

At the two extremes of the exponent $\alpha$ we have two familiar models: {\it (i)} at $\alpha = \infty$ the model reduces to the well-celebrated nearest-neighbor Ising model with transverse field, which can be solved analytically in terms of free fermions~\cite{LIEB1961407}; this model undergoes an equilibrium quantum phase transition 
from ferromagnetic to paramagnetic phase at $h_c/J = 0.5$.
{\it (ii)} At $\alpha = 0.0$ where the model becomes a fully connected, 
and it is also known as Lipkin, Meshkov, and Glick (LMG) model~\cite{LIPKIN1965188,MESHKOV1965199,GLICK1965211}; here
the zero-temperature ferromagnetic and paramagnetic phases are separated by a critical point at $h_c/J = 1$.
In thermodynamic limit, the quantum phase-transition point 
is kept unchanged for all $0\leq \alpha \leq 1$~\cite{Phs_tomo}.

However, it is sufficient to have $\alpha < 2.0 $ to observe a persisting ferromagnetic long range order at low but finite temperature for $|h|$ smaller than a critical value. 
In this regime, the system manifests a plethora of exotic phenomena 
such as non-linear light-cone propagation of correlations~\cite{PhysRevLett.111.207202,PhysRevLett.114.157201,PhysRevA.93.053620}, 
dynamical phase transitions~\cite{dynamical_silva,PhysRevB.100.180402,Jad_LRI,Jad_LRI2}, 
quasi-localization of topological defects~\cite{PhysRevB.99.121112},
and time-crystalline behavior~\cite{PhysRevX.7.011026}.
Recently, pre-thermal stabilization of time-crystalline behavior has been also established for $\alpha > 2.0$~\cite{Pretherm_DTC}.
Some of these phenomena have been investigated 
in a range of experiments with trapped ions~\cite{time_cryst_expt,Dyn_quant_PT_expt_1,Dyn_quant_PT_expt_2,trapp_ion_expt_1,trapp_ion_expt_2}.

\subsection{Full counting statistics}\label{sec:fcs}
The full information on how the order melts after a quench is provided by the time evolution of the full probability distribution function (PDF) of the order parameter. Our order parameter is the longitudinal magnetization defined in a subsystem of length $l$
\begin{equation} \label{eq:fcs1}
    M(l) = \sum_{i=1}^l s^x_i
\end{equation}

Observables that are defined over a fixed subsystem relax locally in space unlike the global observables \cite{Essler_2016} and also have thermodynamic behavior in $l$ unlike localized observables. 
Therefore observables of this kind are suitable choice to study the relaxation dynamics after a global quantum quench. 
The probability that the observable defined in equation~(\ref{eq:fcs1}) will take a value $m$ in a certain state $\ket{\Phi}$ is
\begin{equation} \label{eq:fcs2}
 P_l(m) = \mel{\Phi}{\delta(M(l)-m)}{\Phi} = \int_{-\infty}^{\infty} \frac{d\theta}{2\pi} e^{-i \theta m}\mel{\Phi}{e^{i \theta M(l)}}{\Phi}
\end{equation}
where
$G_l(\theta) = \mel{\Phi}{e^{i \theta M(l)}}{\Phi}$
is the generating function of the moments of $P_l(m)$, and satisfy
the following properties:
$G_l(0) = 1$,
$G_l(-\theta) = G_l(\theta)^*$,
$G_l(\theta+2 \pi) = (-1)^l  G_l(\theta)$.
The first two properties are trivial. 
The last one can be easily verified by exploiting the following identities
\begin{equation}
e^{i \theta M(l)} = \prod_{j=1}^l e^{i \theta S^x_j},
\quad
  e^{i \theta S^x_j}  =
  \cos\bigg(\frac{\theta}{2}\bigg) 
  + i \sin\bigg(\frac{\theta}{2}\bigg) \sigma^x_j  \label{eq:5c} 
\end{equation}
thus implying 
\begin{equation}\label{eq:fcs8}
e^{i (\theta+2\pi) M(l)} = \prod_{j=1}^l \bigg[-\cos\bigg(\frac{\theta}{2}\bigg) - i \sin\bigg(\frac{\theta}{2}\bigg) \sigma^x_j\bigg] 
= (-1)^l e^{i \theta M(l)} .
\end{equation}

Thanks to this periodicity, we can restrict the range of $\theta$ in equation~(\ref{eq:fcs2}) in the interval $-\pi \le \theta < \pi $. Finally, since $m$ can take either integer or half integer values depending upon whether $l$ is even or odd, we can rewrite the PDF as
\begin{equation} \label{eq:fcs9}
P_l(m) = 
     \begin{cases}
       \sum_{r\in \mathbb{Z}} \Tilde{G}_l(r) \delta(m-r) &\quad\text{if }l\text{ is even},\\
        \sum_{r\in \mathbb{Z}} \Tilde{G}_l(r+\frac{1}{2}) \delta(m-r-\frac{1}{2}) &\quad\text{if }l\text{ is odd},\\
     \end{cases}
\end{equation}
where
\begin{equation}\label{eq:fcs10}
\Tilde{G}_l(r) = \int_{-\pi}^{\pi} \frac{d\theta}{2\pi} e^{-i r \theta} G_l(\theta).
\end{equation}

The computational bottleneck of calculating the PDF is the generating function $G_l(\theta)$ after which $P_l(m)$ is obtained by a simple Fourier transformation.

Historically, full counting statistics was introduced in the field of quantum transport in mesoscopic system to study the fluctuation of electron transport \cite{FCS_main}. In quantum transport the moments of electron distribution gives several physical quantities like current, zero-frequency noise and Fano factor. Apart from this FCS has also been used to study entanglement in manybody systems \cite{FCS_ENT_1,FCS_ENT_2}, quantum criticality \cite{FCS_QUANT_CRIT}, and signatures of many body localization \cite{FCS_MBL_1,FCS_MBL_2}.

\subsection{Quench Protocol and numerical details}\label{sec:protocol}
The general quench protocol reads as follow (hereafter $J=1$):
{\it (i)} At $t=0$ the system is initialized in the ground state $\ket{\psi_i}$ of a certain point in the equilibrium phase space denoted by a pair of parameters $(\alpha_i,h_i)$. 
{\it (ii)} The system is then suddenly quenched to a different point in the parameter space, $(\alpha_f,h_f)$, and it is evolved unitarily with the Hamiltonian $H(\alpha_f,h_f)$, accordingly to 
$
\ket{\psi_{t}} = e^{-i t H(\alpha_f,h_f)}\ket{\psi_i}.
$

In order to accomplish the task we have used matrix product state (MPS) based Density Matrix Renormalization Group (DMRG) algorithm~\cite{DMRG1,DMRG2,DMRG_MPS}, 
and we have initialized the system in the ground state of the Hamiltonian $H(\alpha_i,h_i)$.
Notice that, in the case of quenches starting from $h_i = 0$ [{\it cf.} Sec.~\ref{sec:along_h}], the initial state admits an exact MPS representation with $\chi = 2$. Thereafter, for the unitary time evolution we use the Time Dependent Variational Principle (TDVP) algorithm~\cite{TDVP_one,TDVP_two}, with second-order integration scheme, to solve the local forward and backward Schrodinger's equations. In our numerical simulations we used a single-site TDVP recipe and fixed the initial MPS bond dimension; we fixed the Trotter time-step to $dt = 0.05$. There is a finite time step error of $O(dt^3)$ per time step and $O(dt^2)$ per unit time~\cite{Time_MPS}.

 For all cases considered, we have taken the MPS bond dimension $\chi = 100$, which turns out to be largely sufficient to get a representation of the exact ground state with DMRG and the subsequent time evolution with TDVP. In figure \ref{fig:chi_error_1} we plot the relative error of the subsystem magnetization at increasing values of bond dimension, we observe that the error is less than $O(10^{-1})$ overall and less than $O(10^{-3})$ for lower time scales, in which most of our results are based. Furthermore, in figure \ref{fig:EFP_error} we plot the absolute error in Emptiness formation probability $P(m = \frac{\mp l}{2})$ at the two solvable extremes of the model and observe that overall the error remains smaller than $O(10^{-2})$. See Appendix~\ref{app:convergence} for further discussion on this.

We simulated the dynamics of systems with maximum size $L = 200$, and we measured our observable for different subsystems $l$; nevertheless, the majority of the results we presented correspond to the subsystem of size $l = 100$. Let us mention that the bottleneck of the simulation is the local eigensolver in DMRG sweep, and the local exponential solver in TDVP sweep for which we have used Lanczos algorithm~\cite{Lanczos} with full re-orthogonalization\cite{orthogon_lancz}. 
With the MPS representation of the state $\ket{\Phi}$, the generating function $G_l(\theta)$ can be measured by sandwiching the product operator $e^{i \theta M(l)}$ and exploiting usual tensor-network techniques. 
We have chosen the subsystem of size $l$ to be at the center of the full system. Finally, the discrete Fourier transform is used to compute $P_l(m)$, using equation~(\ref{eq:fcs9}) with the range $\theta \in [-\pi,\pi]$ 
and discretization step $d\theta = 0.01$. 

\subsection{Measuring full counting statistics in experiments}

In past years the long range Ising model has been realized in the system of trapped ions with linear radio frequency Paul trap, on various out of equilibrium experiments \cite{info_spread,obs_pretherm,Dyn_quant_PT_expt_1,Dyn_quant_PT_expt_2}, with greater control and tunability. These experimental setup have a high tunability of the interaction range, $0 < \alpha < 3$ \cite{trapped_ion1,quant_PT_expt} and with single shot spin detection method the individual state of each ion spin can be measured with an efficiency of nearly 99 per cent \cite{Dyn_quant_PT_expt_1}. This allows us to directly calculate the subsystem magnetization of the long range Ising model, and by collecting the histogram over several repeated single shot measurements we can get the full probability distribution of our the subsystem magnetization at different time slices after a quantum quench. Such statistical distribution of different parameters have already been measured in several experiments, spin excitation probability distribution in long range Ising model \cite{obs_pretherm}, domain statistics in long range Ising model \cite{Dyn_quant_PT_expt_1}, full counting statistics of time of flight images in one dimensional Bose gas \cite{fcs_TOF}, full contrast distribution of one dimensional Bose gas \cite{full_contrast}.

\section{Results}\label{sec:res}

We investigate the dynamics of the subsystem magnetization PDF after a quantum quench. We consider two different classes of quantum quenches.

\begin{itemize}
    \item We initialize the system at $h_i = 0$ and quench the transverse field to finite $h_f$ for a given value of interaction range.
    
    \item We initialize the system at $\alpha = 0$ (fully connected) and quench the interaction range to $\alpha = 10$ (short range Ising) while fixing the transverse field. We also perform a similar quench in opposite direction. 
\end{itemize}

\subsection{Quench along transverse field}\label{sec:along_h}

\begin{figure}[b!]
    \centering
    \includegraphics[width=1.0\textwidth]{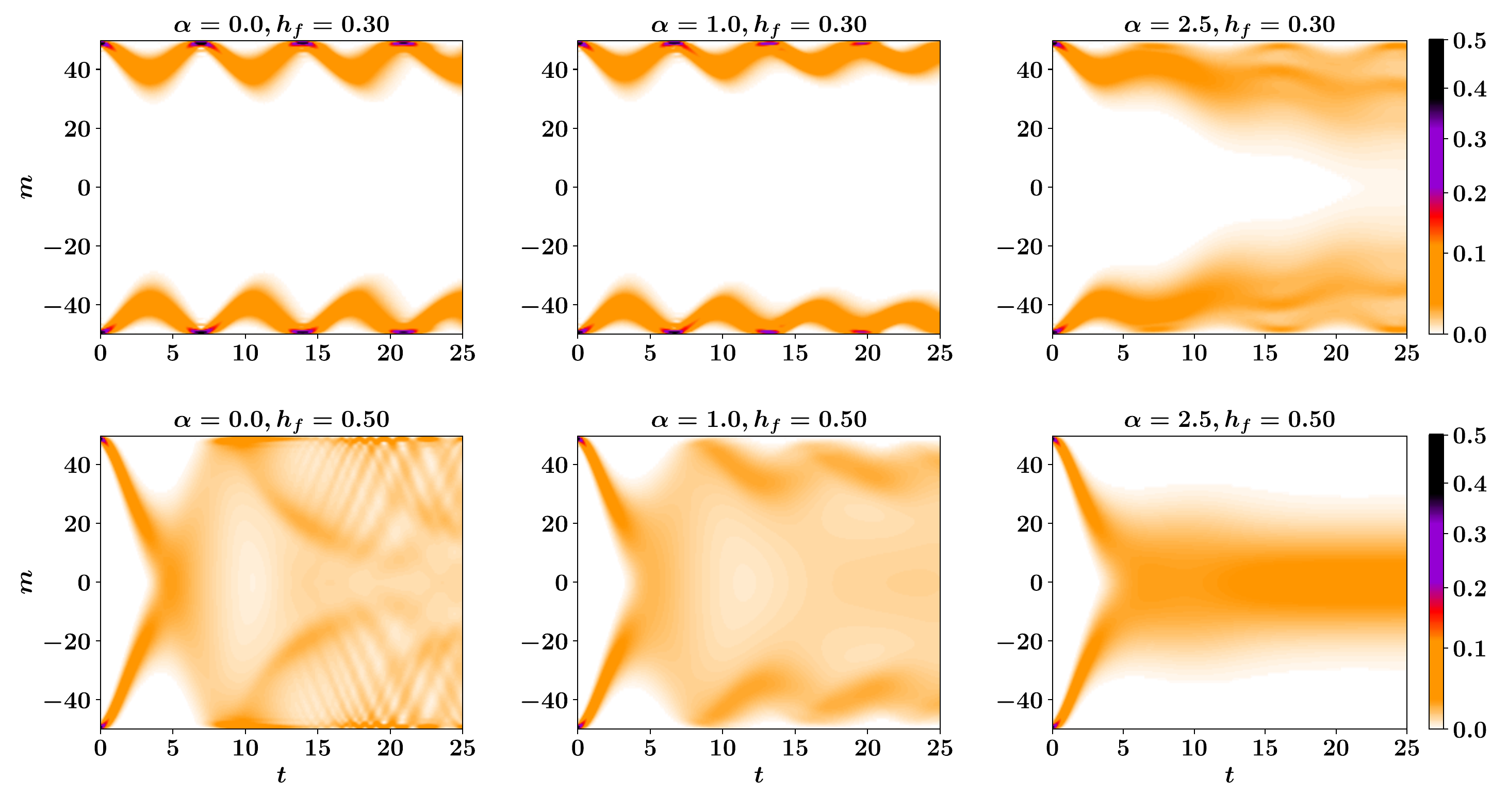}
    \caption{PDF dynamics of subsystem magnetization after a quantum quench for $l=100$, $\alpha \in \{0.0,1.0,2.5\}$ and $h_f \in \{0.30, 0.50\}$ (in first and second rows respectively).}
    \label{fig:first_along_h}
\end{figure}

Here we study the time evolution of the full PDF of the subsystem magnetization $M_l(t)$ after a quantum quench along the direction of transverse field $h$. We will keep our quench entirely within the ferromagnetic region of the ground state equilibrium phase diagram \cite{Ramsey_intferro,Phs_tomo}. We prepare our system in the ground state of the Hamiltonian (\ref{eq:mdl1}) with $h=0$, which is the $\mathbb{Z}_2$ symmetric GHZ state~\cite{GHZ,GHZ_lab}

\begin{equation} \label{eq:GS}
\ket{\psi_i} = \frac{1}{\sqrt{2}}(\ket{\rightarrow,...\rightarrow, \rightarrow, \rightarrow...,\rightarrow}+\ket{\leftarrow,...\leftarrow, \leftarrow, \leftarrow...,\leftarrow}).
\end{equation}

\begin{figure}[b!]
    \centering
    \includegraphics[width=1.0\textwidth]{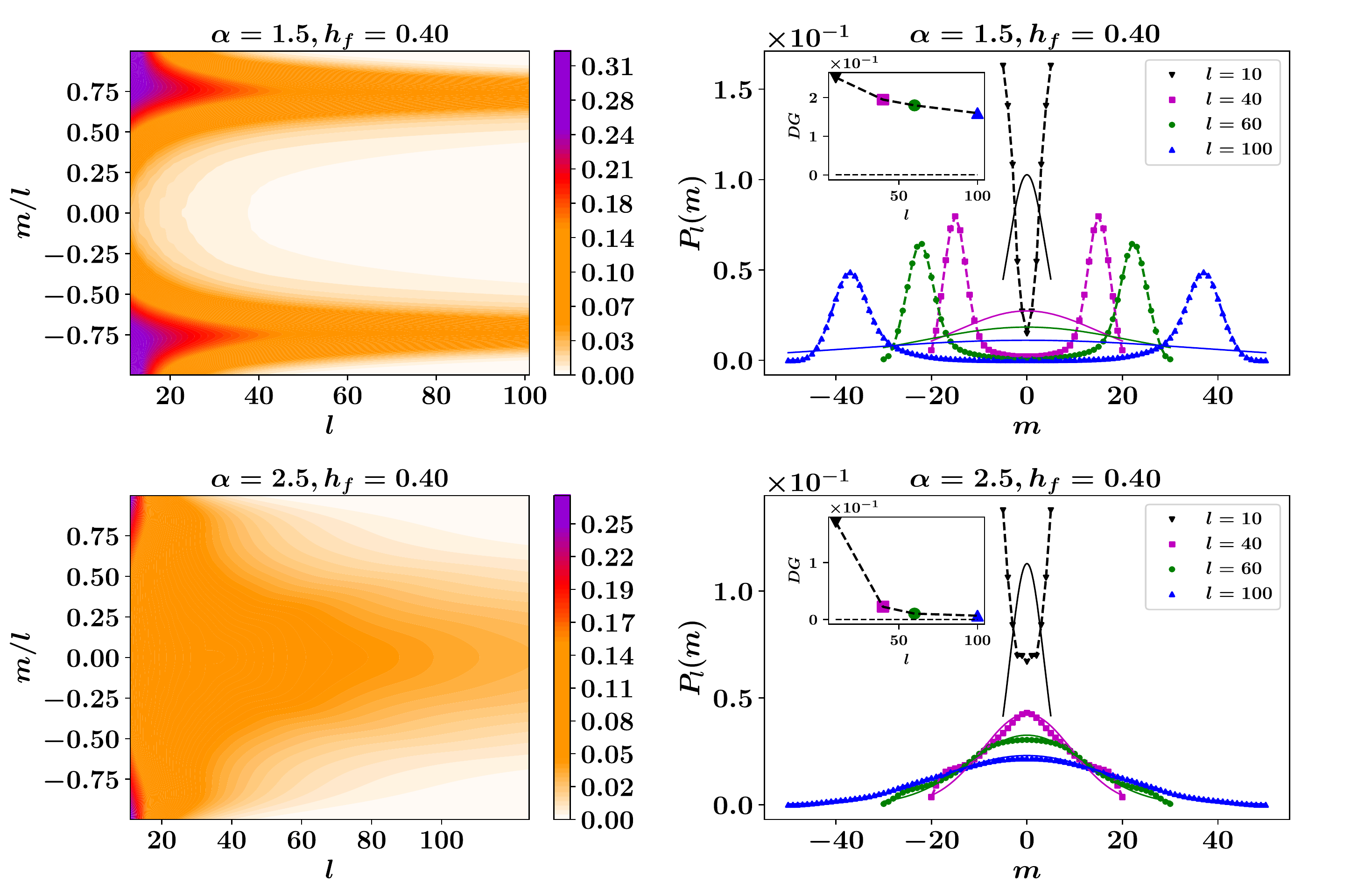}
        \caption{Contour plot of the late time PDF of subsystem magnetization after a quantum quench as a function of subsystem size, for $\alpha \in \{1.5,2.5\} $, $h_f =
        0.40$, and time $t = 25$. The subsystem magnetization $m$ has been rescaled to the range $m/l \in [-1.0,1.0]$, due to this rescaling the intensity of the contour plot decreases with increasing $l$ and the colorbar readings doesn't signify the actual value of PDF. The second column shows PDF at four representative values of $l$. The symbols are TDVP results and the lines are the Gaussian approximation \ref{eq:Gauss}. The inset shows the dependence of $DG$ on $l$.}
        \label{fig:around_2}
\end{figure}

\begin{figure}[b!]
    \centering
    \includegraphics[width=1.0\textwidth]{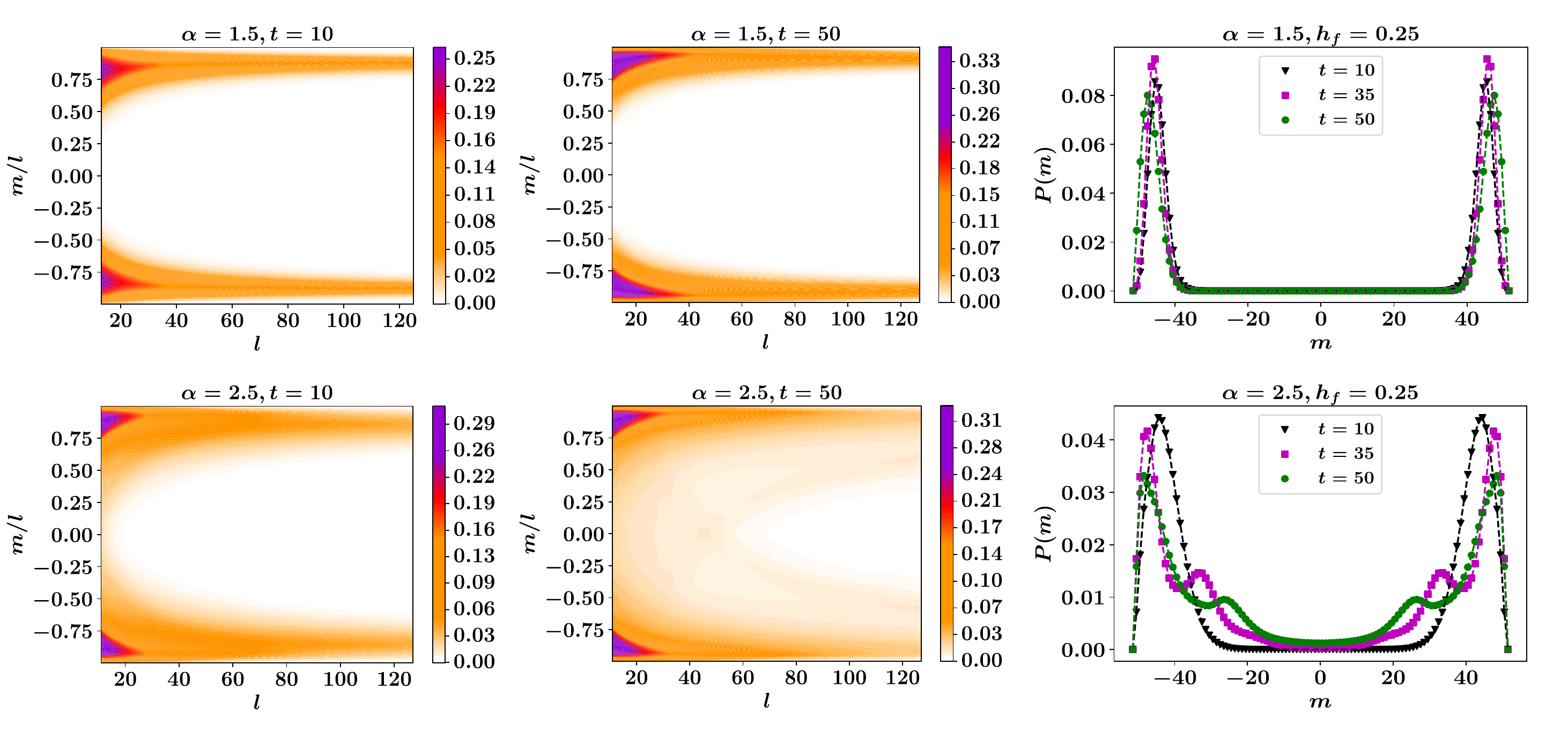}
        \caption{PDF of subsystem magnetization for $\alpha \in \{1.5,2.5\} $, $h_f =
        0.25$ (in first and second rows respectively) at different time scales after the quantum quench. The first two columns shows the contour plot of the rescaled PDF as a function of subsystem size at times $t \in \{10, 50\}$. The last column shows the PDF at times $t \in \{10,35,50\}$ and subsystem size $l=100$.}
        \label{fig:around_2_hf=0.25}
\end{figure}

This state is characterised by a two-peak PDF, $P_l(m) = \frac{1}{2}(\delta_{m,l/2}+\delta_{m,-l/2
})$, thus exhibiting long range ferromagnetic order. We suddenly quench the system to a final hamiltonian $H(h_f)$ with $0 < h_f < h_c(\alpha)$, where $h_c(\alpha)$ is the equilibrium ferromagnetic to paramagnetic transition point at the given $\alpha$. We then evolve the PDF of the subsystem magnetization with post quench hamiltonian in real time with TDVP. The late-time behaviour of the PDF is observed to be strongly dependent on the post quench parameters and the subsystem size $l$. The two extremes of Hamiltonian \ref{eq:mdl1} at $\alpha = \infty$ and $\alpha = 0$ are exactly solvable. In Appendix \ref{app:SHORT_ISING} we compare our numerical results with exact analytical results for short range transverse field Ising model and in Appendix \ref{app:FULL_CONNECT} we provide a recipe for exactly calculating the time dependent PDF for fully connected Ising model with exact diagonalization and compare the TDVP results with the exact results.

In figure \ref{fig:first_along_h} we show this dependence with few representative quenches. In the first row we see that the dynamics is qualitatively the same for $\alpha \in \{0.0,1.0\}$ and $h_f=0.30$. In both cases we see that the system strongly retains the initial long range ferromagnetic order throughout the time evolution. Furthermore, we observe a peculiar oscillation in PDF with a return frequency along the time axis. This behavior completely changes for $\alpha = 2.5$ where the initial ferromagnetic order starts to melts at later time signifying a completely different dynamics. In the second row we observe a different quench dynamics for the same values of $\alpha$, and a larger value of the post-quench transverse field $h_f = 0.50$. Here, the initial ferromagnetic order quickly melts since the larger transverse field works against it. However, depending on the value of interaction range, the time evolution of the PDF undergoes
a qualitative change in its late time dynamical behavior.
For $\alpha = \{0.0,1.0\}$ the dynamics of the PDF is characterised by periodic re-bouncing of probability streams which lead to a broad and flat distribution; for $\alpha = 2.5$ this behavior completely changes with PDF smoothly melting and eventually attaining a Gaussian shape centered around zero.  

\begin{figure}[b!]
    \centering
    \includegraphics[width=1.0\textwidth]{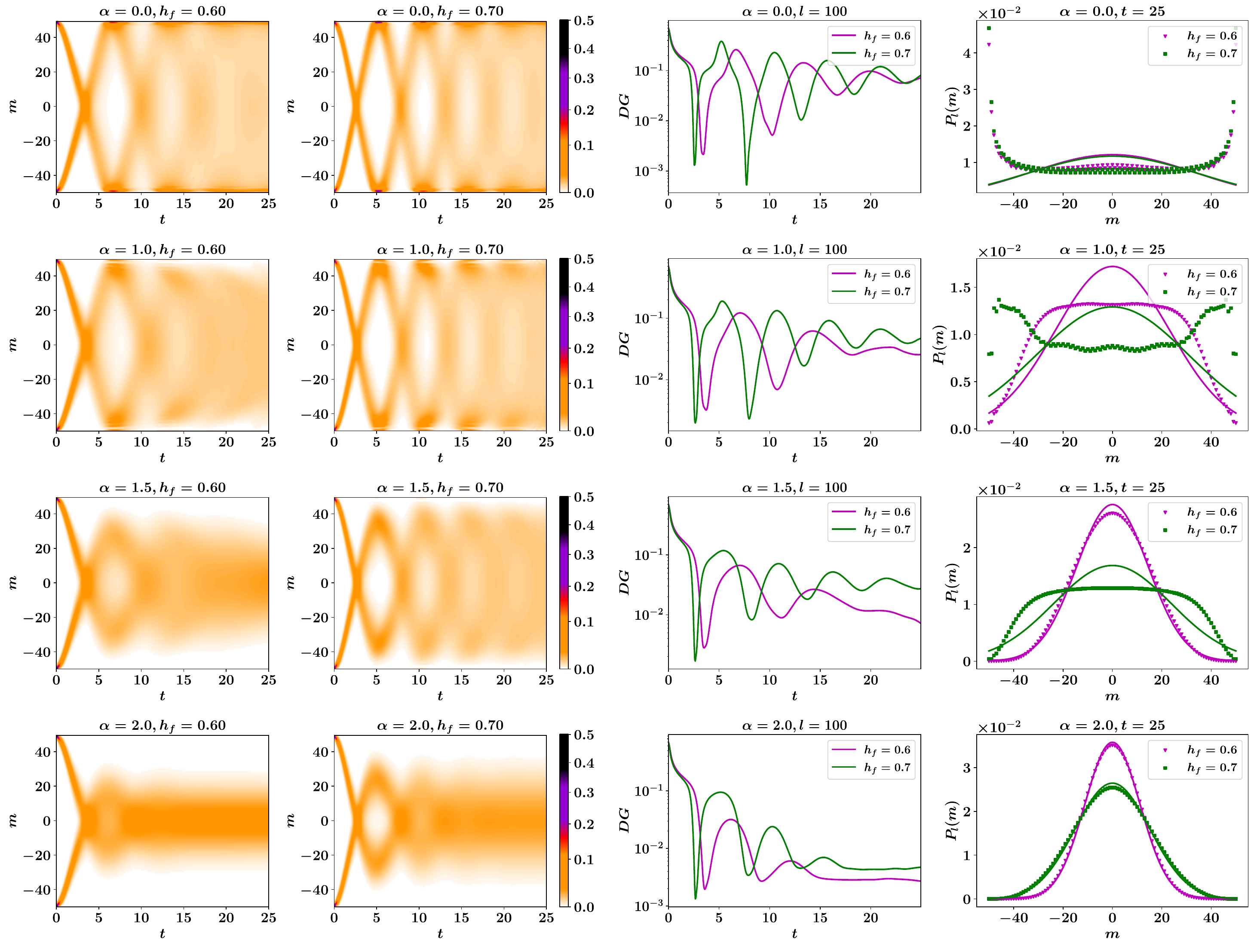}
\caption{PDF of subsystem magnetization after a quantum quench for $l = 100$, $\alpha \in \{0.0,1.0,1.5,2.0\}$ and $h_f \in \{0.60, 0.70\}$. The first two columns shows the evolution of the PDF as a function of time $t \in [0, 25]$. The third column shows the time evolution of DG (y-axis in log scale) and the last column shows the late time PDF at time $t = 25$ for the mentioned parameters. The symbols are the TDVP results whereas the corresponding full lines are the Gaussian approximation \ref{eq:Gauss}.}
\label{fig:quench_above_0.5}
\end{figure}

Gaussification of the PDF is an important behavior because it signifies the complete melting of the long range ferromagnetic order into a paramagnetic one. We expect Gaussification when the linear dimension of the subsystem exceeds the correlation length of the steady state i.e. $l > \xi$. The goodness of Gaussification is measured qualitatively by comparing the PDF with the Gaussian approximation obtained with the first two moments 

\begin{equation} \label{eq:Gauss}
P_l(\mu,t) = \frac{1}{\sqrt{2\pi \sigma^2(t)}}\exp\Bigg[-\frac{(\mu-\Bar{m}(t))^2}{2\sigma^2(t)}\Bigg]
\end{equation}

where $\Bar{m}(t) = \bra{\psi_t}M(l)\ket{\psi_t}$ and $\sigma(t) = \bra{\psi_t}(M(l)-\Bar{m}(t))^2\ket{\psi_t}$ are the first two moments of our order parameter. Quantitatively the goodness of Gaussification can be measured by defining a metric Distance to
Gaussian (DG) as

\begin{equation} \label{eq:dist_Gauss}
DG = \sqrt{\sum_{m}[P(m)-P_G(m)]^2}
\end{equation}

where $P(m)$ is the PDF calculated numerically and $P_G(m)$ is the corresponding Gaussian PDF approximated using \ref{eq:Gauss}. DG is the measure of how close (or far) is the PDF from the Gaussian shape, DG = 0 implies a perfect Gaussian shape. This measure is used in \cite{DG} under the name Distance to Thermalization (DT). $DG$ might not be a proper metric for cases in which the system does not relax to a steady state in the given time frame and shows oscillations, in such cases we introduce the time averaged $DG$ as

\begin{equation} \label{eq:dist_Gauss_T_avg}
DG_{avg} = \frac{1}{T - T_o} \int_{T_o}^{T} DG(dt) dt
\end{equation}

where $T_o$ is chosen to avoid the initial sharp drop in $DG$ [{\it cf.} Fig.~\ref{fig:quench_above_0.5}].

In figure \ref{fig:around_2} we plot the dependence of late time PDF of order parameter on subsystem size $l$ for two representative values of interaction range above and below $\alpha = 2.0$ and $h_f = 0.40$. We observe two characteristically different behavior of PDF in these two regimes. For $\alpha = 1.5$ the PDF remains double peaked for all values of $l$ and further, the two branches of PDF diverges with increasing $l$ suggesting that in thermodynamic limit the initial memory of long range order is strongly retained. For $\alpha = 2.5$ the PDF is double peaked for smaller $l$, becomes flat  for intermediate $l$ and eventually becomes Gaussian for large $l$, suggesting that in thermodynamic limit the initial memory of long range order completely melts. These two characteristically different behavior of PDF above and below $\alpha = 2.0$ is the basis of dynamical quantum phase transition based on order parameter (DQPT-OP) as proposed in \cite{dynamical_silva} according to which $h_f \lesssim 0.50$, $\alpha = 2.0$ marks the transition line between dynamical ferromagnet and dynamical paramagnet.

Alternatively, it has been argued in \cite{Jad_LRI} that DQPT-OP, based on prethermal values of order parameter, persists well beyond $\alpha = 2.0$ and is dependent on both initial and final quench parameters. In figure \ref{fig:around_2_hf=0.25} we performed quenches similar to figure \ref{fig:around_2} at a smaller value of final transverse field, $h_f = 0.25$ and for a longer timescale $t = 50$. At $\alpha = 1.5$ we observe a qualitatively similar behavior at all timescales and subsystem sizes, the initial ferromagnetic order persists throughout the time evolution after a quantum quench and the PDF maintains a distinct double peak at all times. At $\alpha = 2.5$ we observe a completely different behavior, while we observe long range ferromagnetic order at $t = 10$ for all the simulated system sizes, at $t=50$ the long range order start to melt. This behavior is more clearly observed in the rightmost panel of the second row of figure \ref{fig:around_2_hf=0.25} where the PDF for $t = 10$ shows a distinct double peak which becomes less and less prominent with increasing time, together with a stream of probability density travelling toward $m=0$.
Based on these observations we can argue that for a fairly bigger system size and longer timescale we will observe a complete meltdown of the initial long range ferromagnetic order and most possibly Gaussification of the PDF in this regime. This translates to the lack of DQPT-OP beyond $\alpha = 2.0$.

\begin{figure}[b!]
    \centering
    \includegraphics[width=\linewidth]{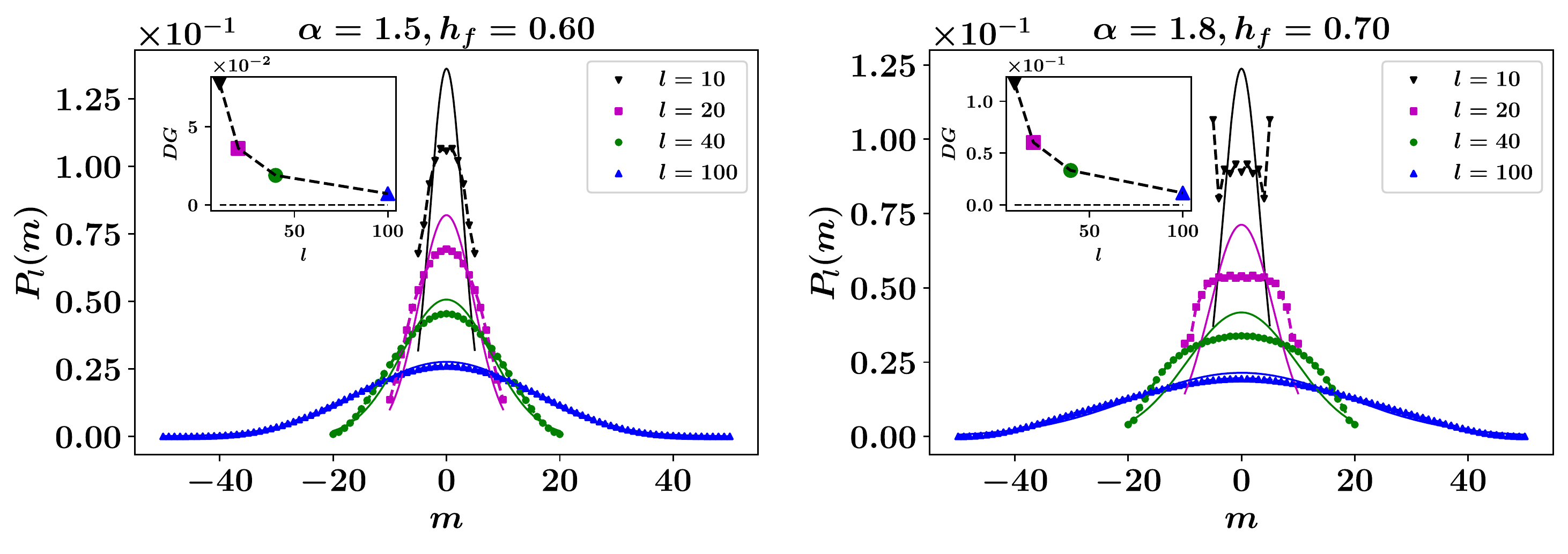}
\caption{The PDF of subsystem magnetization after a quantum quench at time $t = 25$ for subsystem sizes $l \in \{10,20,40,100\}$ for two representative values of post quench parameters $\alpha = 1.5, h_f = 0.60$, and $\alpha = 1.8, hf = 0.70$. The symbols are the TDVP results whereas the corresponding full lines are the Gaussian approximation \ref{eq:Gauss}. In the inset we show the dependence of DG on the subsystem size.}
\label{fig:finite_size}
\end{figure}

In figure \ref{fig:quench_above_0.5} we explore the quench dynamics of PDF for $h_f > 0.50$ at four different representative values of $\alpha$. Quenches in this region provides us with interesting ideas on the dependence of PDF dynamics on the post quench parameters and the subsystem size. We see that for the fully connected case of $\alpha = 0.0$ the initial  ferromagnetic order melts and after few oscillation (which depends on $h_f$) broadens in shape. The time evolution of the metric DG follows a similar pattern and at late times it oscillates and stays well above the zero signifying no Gaussification. The system however retains some memory of initial ferromagnetic order as can be seen from the PDF at $t=25$ that shows two peaks at the edges for both values of $h_f$. At $\alpha = 1.0$ we see two markedly different behavior of PDF at $h_f = 0.60$ and $0.70$. At $h_f = 0.60$ although the late times PDF is flat and far from Gaussian we don't observe any remnants of initial ferromagnetic order while at $h_f = 0.70$ we still observe some remnants of initial ferromagnetic order as shown by the peaks at the edges of late time PDF. These two different behaviors shows the strong dependence of dynamics of PDF after quantum quench on the depth of the quench. As the depth of the quench is increased more energy is injected into the system due to which the system takes longer time to relax to a steady state and consequently we observe more oscillations in the evolution of PDF. At $\alpha = 1.5$ we observe Gaussification of PDF at late times for $h_f = 0.60$, however on increasing the quench depth to $h_f = 0.70$ the PDF becomes a flat for the same simulation time. In this case the time evolution of the metric DG provides a clearer picture as for $h_f=0.60$ DG starts to relax to zero whereas for $h_f = 0.7$ it oscillates and stays well above zero. Further increasing the interaction range to $\alpha = 2.0$ we observe clear Gaussification at both values of $h_f$. The time evolution of DG also shows that the system relaxes to a stationary state faster and to a lower value for $h_f = 0.60$ than $h_f = 0.70$.\par 

\begin{figure}[b!]
    \centering
    \includegraphics[width=\linewidth]{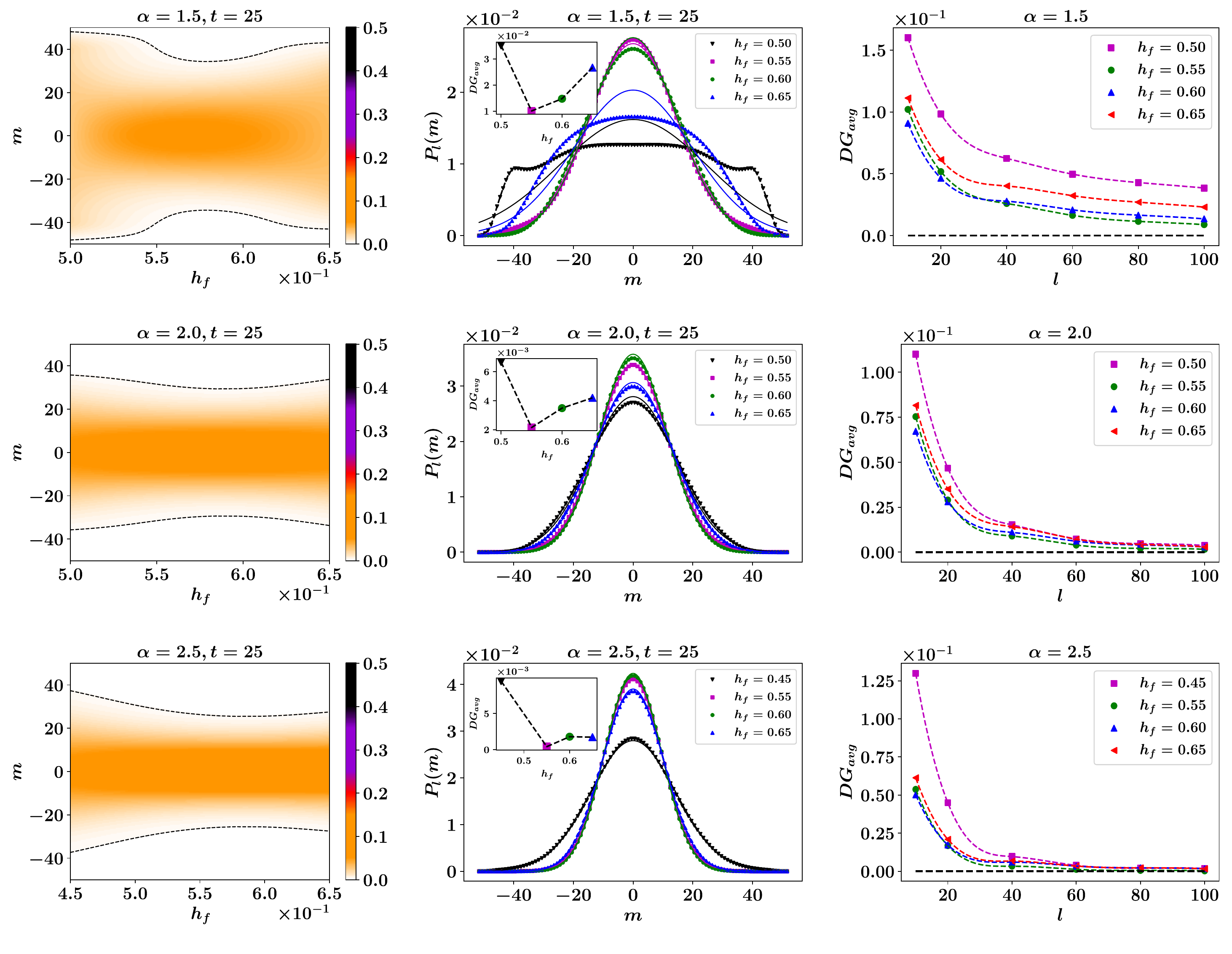}
\caption{PDF dynamics of subsystem magnetization after a quantum quench as a function of post quench transverse field $h_f$ at three representative values of interaction range $\alpha = \{1.5, 2.0, 2.5\}$. The dotted lines are the lines of constant probability at $P(m) = 0.002$, these lines don't have any quantitative significance and is plotted for better visualization of PDF. The first column shows the color plot of PDF at time $t = 25$ at different values of $h_f$ and $l = 100$. The second column shows the PDF at $t = 25$ at different $h_f$, the symbols are the TDVP calculations whereas the lines are the Gaussian approximation \ref{eq:Gauss}. The inset shows $DG_{avg}$ at corresponding values of $h_f$. The last column shows the finite size dependence of $DG_{avg}$ for four different values of $h_f$. $DG_{avg}$ is calculated by averaging over the final $\Big(\frac{3}{5}\Big)^{th}$ of the total simulation time.}

\label{fig:diff_h_f}
\end{figure}

In figure \ref{fig:quench_above_0.5} we observed Gaussification only for higher values of $\alpha$. For the simulation time and subsystem size available to us we observed mostly flat distribution for smaller values of $\alpha$. Solely based on these observations we cannot make any claim about the Gaussification of the PDF of our order parameter. However, studying the PDF for several values of subsystem size $l$ for selected post quench parameters allow us to make a stronger claim about the Gaussification of PDF in the thermodynamic limit. In figure \ref{fig:finite_size} we show the behavior of the PDF for two representative quenches for different subsystem sizes at $t = 25$. For $l =100$ we see Gaussification for both quenches, however for smaller subsystem sizes we observe mostly flat distribution far from Gaussian. The variation of DG with system size in the inset shows the approach to Gaussian with increasing system size. These flat distributions are qualitative similar to the ones we saw in figure \ref{fig:quench_above_0.5} for smaller $\alpha$ but with $l=100$. For $\alpha =1.8, h_f=0.70, l=10$  we even observe a strong remnants of initial ferromagnetic long range order shown by two sharp peaks at the edges. Based on these observations we can argue that the flat distribution is an intermediate distribution between the double peaked and Gaussian distribution. For a sufficiently large subsystem size and long simulation time, we will observe Gaussification for smaller values of $\alpha$. This observation is in line with DQPT-OP proposed in \cite{dynamical_silva} which suggests the dynamical critical point to be close to $h_f \approx 0.50$. The exact values of the dynamical critical points and the critical exponents are still an open question. This phenomenon has been experimentally observed with trapped ion qubits~\cite{Dyn_quant_PT_expt_1}.

In figure \ref{fig:diff_h_f} we explore the dependence of Gaussification on the post quench transverse field $h_f$ at three representative values of interaction range $\alpha = \{1.5, 2.0, 2.5\}$. For $\alpha = 1.5$ we observe that the system moves towards Gaussian with increasing $h_f$ up to a certain point which is marked by the dip in the color plot PDF versus $h_f$ in the first panel of the first row. On increasing $h_f$ further the PDF moves away from Gaussian and broadens in shape. This is observed more clearly in the second panel of the first row where we observe Gaussification only for the intermediate values $h_f = \{0.55, 0.60\}$. Although higher transverse field tend to destroy the long range ferromagnetic order, we also inject a larger amount of energy in the system as we increase the depth of the quench i.e. $h_f$. This means a higher effective temperature \cite{Essler_2016} and consequently the system requires a longer time and bigger subsystem size to eventually relax to a Gaussian. The intermediate region is a region of compromise where we observe a distinct Gaussian behavior of the PDF of our order parameter. The third panel shows that for sufficiently large system size the PDF relax more and more towards Gaussian for all values of $h_f$. For $\alpha \geq 2.0$ the PDF is Gaussian in the range of $h_f$ and simulation parameters considered, however we still see the dip in the color plot signifying the region of better Gaussification. This dip becomes less and less prominent with increasing $\alpha$ as shown by the plots for $\alpha = 2.5$ in the third row.

\subsection{Quench of the interaction range}\label{sec:along_a}

\begin{figure}[b!]
    \centering
    \includegraphics[width=\textwidth]{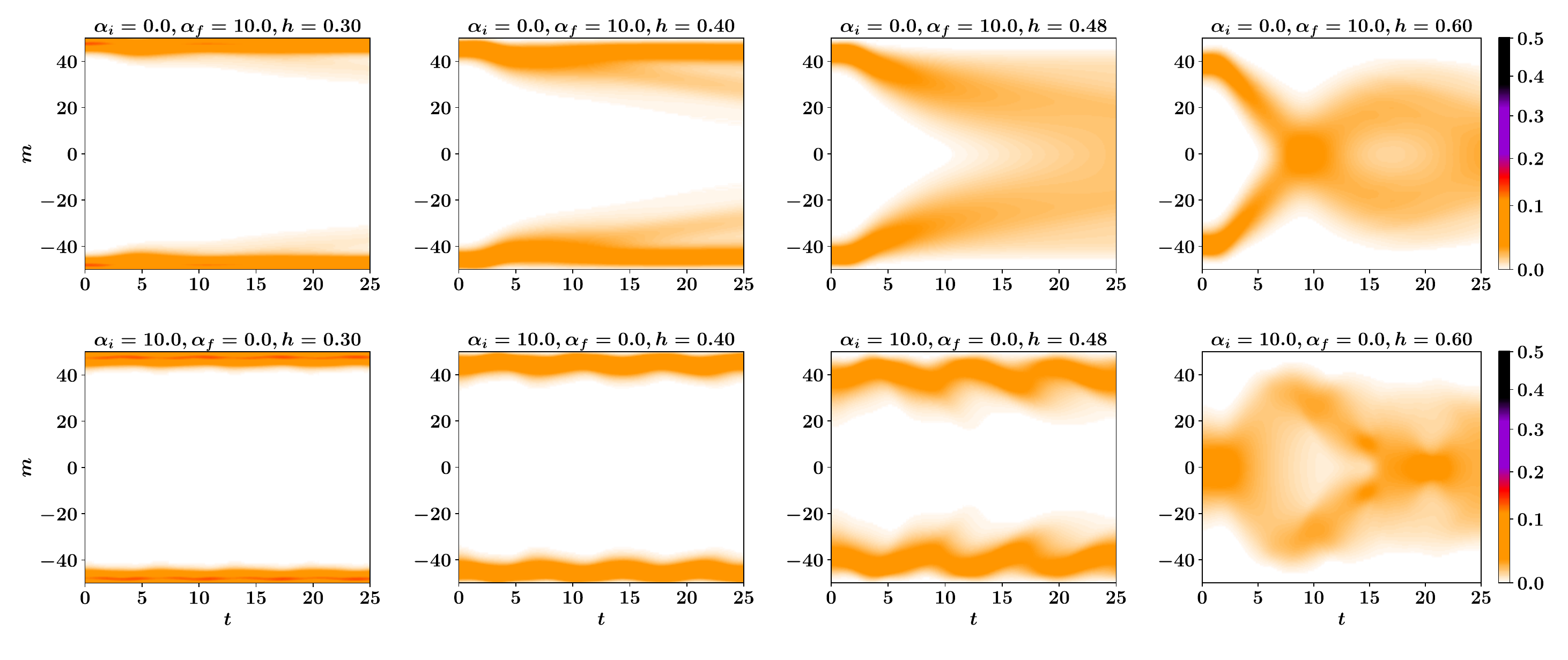}
    \caption{PDF dynamics of subsystem magnetization after a quantum quench along the direction of interaction range, $\alpha$. The first row shows the results for quenches where the system is initialized at $\alpha_i = 0.0$ and quenched to a final $\alpha_f = 10.0$, the second row shows the results for opposite condition. The quenches are performed for three representative values transverse field $h \in \{0.30,0.40,0.48,0.60\}$.}
    \label{fig:along_a}
\end{figure}

Here we study the relaxation of the PDF of the subsystem magnetization $M_l(t)$ following a quantum quench in the direction of interaction range $\alpha$. In particular we quench the system from one extreme to the other; initializing at $\alpha = 0.0$, which is the fully connected model, and quenching to $\alpha = 10.0$ which is almost the Ising model with nearest neighbor interactions, and vice versa. The transverse field $h$ is kept constant throughout the evolution. Apart from looking how and if the initial ferromagnetic order melts at late time, we are also interested to see if quenches in opposite directions are qualitatively equivalent. In Figure \ref{fig:along_a} we show these quenches for four representative values of the transverse field $h = \{0.30,0.40,0.48,0.60\}$. The first thing we immediately notice is that the two peaks of the initial PDF are broader compared to the cases where the system is initialized with $h = 0$. This is due to the fact that now the system has been initialized closer to the equilibrium phase transition line where the PDF peaks are broader and closer to each other.
Remarkably, we do observe a completely different behavior for the two different class of quenches. We see that for quenches starting from the fully connected state and evolving according to the short-range Hamiltonian, the initial ferromagnetic order melts for sufficiently large value of the transverse field. 
For $h = \{0.30,0.40\}$ we observe that the two branches of the PDF start to melt with time, as shown by the stream of probability density branching out from the main PDF peak, although the initial ferromagnetic order effectively remains throughout the evolution. For $h = 0.48$ we observe the complete meltdown of the initial ferromagnetic order with some hints of Gaussification at late times. On the other hand, for quenches starting from the ground state of the short-range Hamiltonian and evolving according to the fully connected Hamiltonian, we observe strong remnants of the initial ferromagnetic order throughout the evolution with no sign of meltdown. We see that although the PDF becomes more oscillatory with increasing $h$ there is no change in the intensity of the PDF branches with time. The last column shows the result for similar quenches but at $h = 0.6$, these quenches are different from others in the sense that the point $\alpha = 10.0, h = 0.6$ lies in the paramagnetic regime of the equilibrium phase diagram. For quench from $\alpha = 0.0 $ to $\alpha = 10.0$ we observe that the initial ferromagnetic order melts faster than before now that the system is quenched to a point in paramagnetic regime in equilibrium phase diagram. The quench in opposite direction is initialized at a paramagnetic point so the PDF begins as a Gaussian which relaxes quickly. Interestingly, at intermediate times we observe an appearance of double peak PDF signifying long range ferromagnetic order. This long range order is short lived and melts quickly. However, the fate of the PDF in a long time limit is not very clear.

The two cases of $\alpha$ considered here are two extremes of the  long range Ising model. When $\alpha = 10.0$ the system is close to well known transverse field Ising model. This model does not support any long range order at finite temperature, thus it is not surprising to observe a melting of the initial ferromagnetic order after the quench; indeed, the protocol injects a finite amount of energy to the system, simulating a finite temperature environment. So while we quench the system from one point in ferromagnetic region to another point in ferromagnetic region in zero temperature equilibrium phase diagram, the system actually relax to a paramagnetic point in a finite temperature phase diagram. On the other hand, the fully connected Ising model supports long range ferromagnetic order even at finite temperature \cite{Phs_tran_gt_alp_2_1,Phs_tran_gt_alp_2_2,gonzalez_finitetemperature} which is presumably the reason why we observe strong remnants of the initial ferromagnetic order at late time.  

\section{Conclusion and Outlook}\label{sec:conclude}
We studied the dynamics of the PDF of the subsystem magnetization in the long range Ising model after a quantum quench. We constrained most of our quenches within the ferromagnetic region of the equilibrium phase diagram as we expect non-trivial dynamics in this region. We studied quenches along the transverse field and interaction range. We found that the dynamics of the order parameter strongly depends on the post quench parameters. Based on these observations we showed that for $\alpha > 2.0$ the initial long range ferromagnetic order eventually melts following a quantum quench in the direction of transverse field. This is signified by the melting of the double peak PDF into a Gaussian one in the long time limit and for large system sizes. For $\alpha \leq 2.0$ however we see a strong remnant of the initial long range order
for quenches in the transverse field such that $h_f \lesssim 0.50$, where a double peak structure in the PDF remains throughout the time evolution. In the region above $h_f = 0.50$ we observed Gaussification of the order parameter PDF for increasing value of $\alpha$. Gaussification of the order parameter PDF in this region is dependent on the size of the subsystem (and eventually the size of the system we can simulate) and the total simulation time, which greatly constrained our numerical work. However, with a finite size analysis for some representative quenches, we can safely claim that for sufficiently large system size and longer simulation time we expect Gaussification of the order parameter PDF for all $\alpha$ following a quantum quench along the transverse field with $h_f \geq 0.50 $.

For quenches along the interaction range, we found qualitatively different dynamics of the order parameter PDF, depending on the direction of the quench. While for quenches starting form the fully connected state and evolving with short-range Hamiltonians we saw an effective melting of the initial ferromagnetic order, we observed a complete persistence of the initial order for quenches in the opposite direction. 

Attaining Gaussification of the order parameter PDF after quenches from one point in the ferromagnetic region of the zero temperature equilibrium phase diagram to another point within the ferromagnetic region is a non trivial phenomenon, suggesting that the system relaxes to a paramagnetic point in the finite temperature phase diagram. 
These results open up the possibility to further study the thermalization dynamics in the long range Ising model. In fact, the non-equilibrium results could be compared with a finite temperature analysis of the model, by observing if the late-time subsystem magnetization after the quench converges to the thermal expectation value corresponding to the effective temperature fixed by the initial value of the energy density~\cite{Essler_2016}. So far, mainly short range models have been investigated in this perspective~\cite{order_melt,dyn_con}; for long range models a clear pathway is to combine the present analysis with matrix product density operator(MPDO)~\cite{MPDO} based TDVP, modified to generate finite temperature states starting from the maximally mixed infinite temperature density matrix~\cite{purify_open,one_D_open}.

\section*{Acknowledgements}
We are thankful to Fabian Essler, Adriano Angelone, Eduardo Gonzalez Lazo for reading the draft and providing valuable comments. Nishan Ranabhat thanks Guglielmo Lami for helpful discussion of the exact solution of fully connected Ising model. The numerical simulation of this project was performed at the Ulysses v2 cluster at SISSA.

\begin{appendix}

\section{Comparing the large $\alpha$ results with exact analytical results for nearest neighbor transverse field Ising} \label{app:SHORT_ISING}

\begin{figure}[h]
    \centering 
  \includegraphics[width=0.9\linewidth]{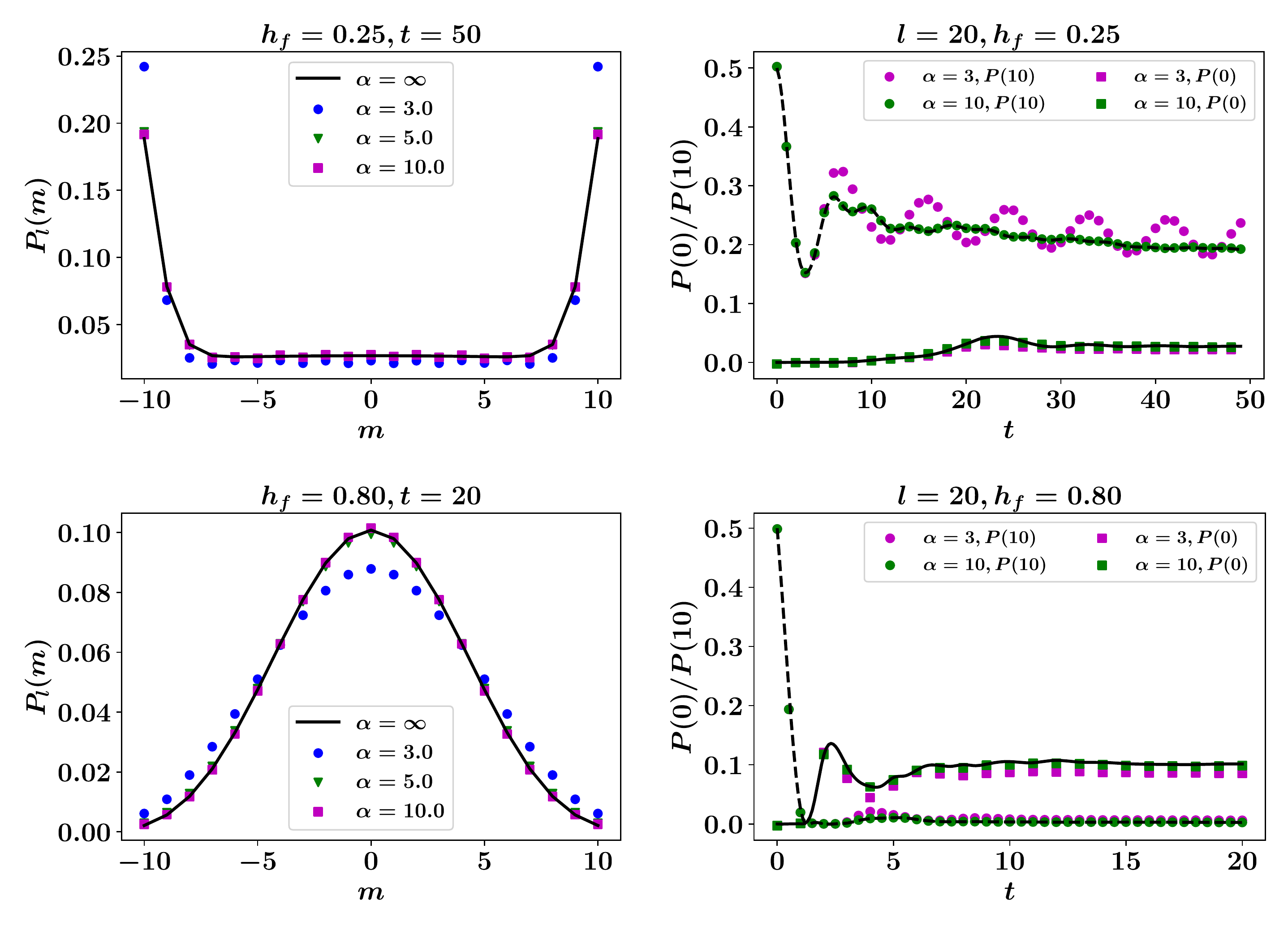}
\caption{PDF of subsystem magnetization of nearest neighbor transverse field Ising model after quantum quench. The first panels on the left column are the PDF after quench to two representative final transverse field and at three increasing values of $\alpha$. The panels on the right are the time evolution of formation probabilities, $P_l(m = \frac{\pm l}{2}), P_l(m = 0)$, for the corresponding final transverse fields. The lines are analytical results \cite{relax} whereas the symbols are TDVP results.}
\label{fig:short_com}
\end{figure}


The long range Ising model~\ref{eq:mdl1} at $\alpha = \infty$ is exactly solvable;
it reduces to the well celebrated nearest neighbor transverse field Ising model. 
The dynamic and stationary PDF for nearest neighbor Ising model has been analytically computed for quenches to both ferromagnetic and paramagnetic regimes by means of a relation to a 3-state classical model \cite{relax} . Here, we compare the PDF after a quantum quench in long range model with increasing values of $\alpha$ obtained by TDVP with the stationary PDF of the nearest neighbor transverse field Ising model computed analytically. We also compare the corresponding evolution of the formation probabilities $P_l(m = \frac{\mp l}{2})$ and $P_l(m=0)$ for quenches to ferromagnetic and paramagnetic regimes with the analytical results.

In figure \ref{fig:short_com} we see that the TDVP results perfectly overlaps with the analytical results for sufficiently large  $\alpha$. In fact we don't observe much difference between the $\alpha = 5$ and $\alpha = 10$ results.

\section{Exact solution for fully connected Ising model}\label{app:FULL_CONNECT}

The other extreme of long range Ising model at $\alpha = 0$ gives us the fully connected Ising Hamiltonian
\begin{subequations}
\label{eq:mdl_fc}
\begin{align}
    H(h) &= -\frac{1}{N}\sum_{i,j}^N \hat{s}^x_i \hat{s}^x_j - h \sum_{i=1}^N \hat{s}^z_i\\
    &= -\frac{1}{N} (\hat{S}^x)^2 - h S^z
\end{align}
\end{subequations}

where $\hat{S}^a = \sum_i^N \hat{s}_i^a$, $a = x,y,z$, are the collective spin operators and has commutation relations like normal spin operators. In this regime the model behaves as a single collective spin and we can exactly solve the dynamics of our initial state with this Hamiltonian. Our initial state is the $\mathbb{Z}_2$ symmetric GHZ state

\begin{equation} \label{eq:GS_fc}
\ket{\psi_0} = \frac{1}{\sqrt{2}}(\ket{\rightarrow,...\rightarrow, \rightarrow, \rightarrow...,\rightarrow}+\ket{\leftarrow,...\leftarrow, \leftarrow, \leftarrow...,\leftarrow}).
\end{equation}

\begin{figure}[ht]
    \centering 
  \includegraphics[width=0.9\linewidth]{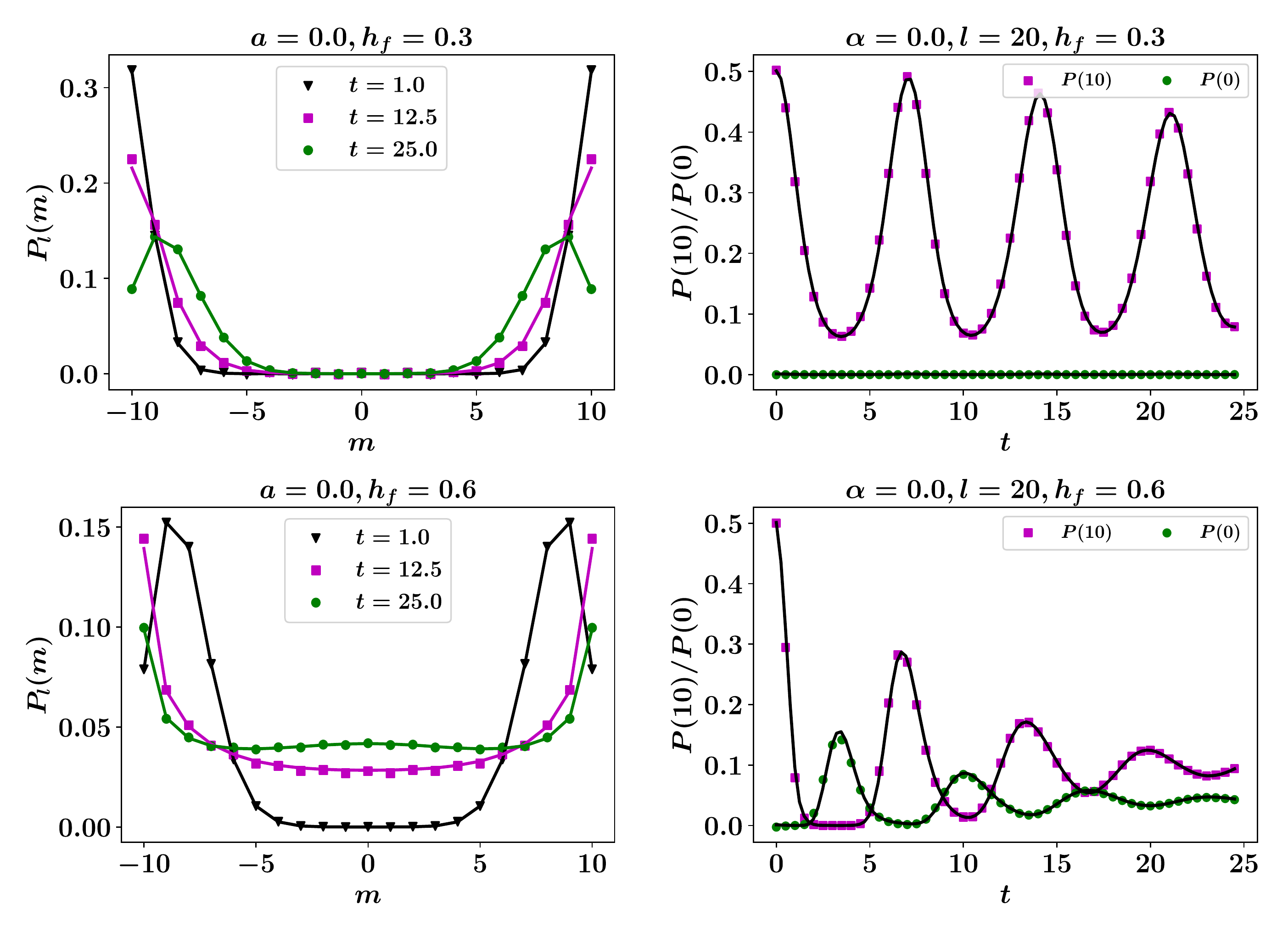}
\caption{PDF of subsystem magnetization of fully connected Ising model after quantum quench. The first panels on the left column are PDF at three representative time slices at two values of final transverse field. The panels on the right are the time evolution of formation probabilities, $P_l(m = \frac{\pm l}{2}), P_l(m = 0)$, for the respective final transverse fields. The full lines are exact results calculated from equation \ref{eq:gf_fin} whereas the symbols are TDVP results.}
\label{fig:fulcon_com}
\end{figure}

This state evolves unitarily as $\ket{\psi_t} = e^{-i H t}\ket{\psi_0}$, so we need to represent $\ket{\psi_0}$ as the superposition of the eigenkets of the Hamiltonian \ref{eq:mdl_fc}. This Hamiltonian is diagonalized with the basis $\big\{\ket{\frac{N}{2},n}\big\}$ that are the simultaneous eigenkets of collective spin operators $(\hat{S}^2,\hat{S}^x)$ in the maximal total spin sector. We strictly remain in this sector because our initial state $\ket{\psi_0}$ is in this sector and the unitary operation $e^{-i H t}$ preserves the total spin quantum number. The total spin quantum number is fixed through out the time evolution and is thus redundant, so we will from hereon use the notation $\{\ket{n}\}$ where $n$ is the number of down spins such that 

\begin{equation}\label{eq:basis}
\ket{n} = \frac{1}{\sqrt{\binom{N}{n}}} \sum_{j_1<j_2<...j_n} \ket{...j_1..j_2...j_n...},
\quad
\hat{S}^x\ket{n} = \Bigg(\frac{N}{2} - n \Bigg) \ket{n}
\end{equation}

where $j_1,j_2,...j_n$ are the positions of the down spins and the sum runs over all $\binom{N}{n}$ configurations. We can represent our initial state \ref{eq:GS_fc} as the linear combination of the eigenkets $\{\ket{E_i}\}$ of the Hamiltonian \ref{eq:mdl_fc}

\begin{equation}\label{eq:ini_eigH}
\ket{\psi_0} = \sum_{j=0}^N k_j \ket{E_j}
\end{equation}

where, $\big\{\ket{E_j}\big\}$ can be represented as the linear combination of the basis states, $\ket{E_j} = \sum_{n=0}^N c_n^j \ket{n}$. The coefficients $\{k_j\}$ can be extracted as

\begin{equation}\label{eq:coeff}
    k_j = \braket{E_j}{\psi_0}
    = \sum_{n=0}^N (c_n^j)^* \frac{\delta_{0,n}+\delta_{N,n}}{\sqrt{2}}
    = \frac{(c_0^j)^* + (c_N^j)^*}{\sqrt{2}}.
\end{equation}

The evolved state is 

\begin{equation}\label{eq:psi_t}
        \ket{\psi_t} = e^{-i H(h) t} \sum_{j=0}^N k_j \ket{E_j}
        = \sum_{j=0}^N k_j e^{-i E_j t} \ket{E_j}
        = \sum_{n=0}^N X_n(t) \ket{n}
\end{equation}

where the time dependent coefficient is introduced as

\begin{equation}\label{eq:t_coeff}
X_n(t) = \sum_{j=0}^N k_j c_n^j e^{-i E_j t}.
\end{equation}

The generating function of the PDF can then be calculated as

\begin{equation}\label{eq:gen_fun}
G_l(\theta,t) = \bra{\psi_t}e^{i \theta \hat{M}(l)} \ket{\psi_t}
= \sum_{n,\Tilde{n}}X_{\Tilde{n}}(t)^* X_{n}(t) \bra{\Tilde{n}}e^{i \theta \hat{M}(l)}\ket{n}.
\end{equation}

Since $\ket{n}$ is not an eigenket of the subsystem magnetization we have to decompose it as

\begin{equation}\label{eq:basis_decom}
    \ket{n} = \frac{1}{\sqrt{\binom{N}{n}}} \sum_{\sigma =max(0,n-N+l)}^{min(l,n)} \ket{\sigma}_l \otimes \ket{n-\sigma}_{N-l}
\end{equation}

where $\sigma$ is the number of down spins in the subsystem of size $l$ such that

\begin{equation}\label{eq:decom_unit}
    \ket{\sigma}_l \otimes \ket{n-\sigma}_{N-l} =
    \sum_{\substack{j_1<j_2<...j_{\sigma} \\ \Tilde{j}_1<\Tilde{j}_2<...\Tilde{j}_{n-\sigma}}}
   \ket{...j_1..j_2...j_{\sigma}...}_l\otimes \ket{...\Tilde{j}_1..\Tilde{j}_2...\Tilde{j}_{n-\sigma}...}_{N-l}
\end{equation}

where $j_1,j_2,...j_{\sigma}$ are the positions of down spins in the subsystem $l$ and $\Tilde{j}_1,\Tilde{j}_2,...\Tilde{j}_{n-\sigma}$ are the positions of down spins in rest of the system. The sum runs through all $\binom{l}{\sigma} \times \binom{N-l}{n-\sigma}$ configurations. The state in equation \ref{eq:basis_decom} is properly normalized as can be seen from the following identity,

\begin{equation}\label{eq:chu_van}
\sum_{\sigma =max(0,n-N+l)}^{min(l,n)} \binom{l}{\sigma} \times \binom{N-l}{n-\sigma} = \binom{N}{n}.
\end{equation}

It goes by the name Chu-Vandermonde identity \cite{chu_Van}.
The representation in \ref{eq:basis_decom} is particularly useful because $\ket{\sigma}_l \otimes \ket{n-\sigma}_{N-l}$ is an eigenket of the of the subsystem magnetization operator

\begin{equation}\label{eq:sub_mag_ket}
    \hat{M}(l)\ket{\sigma}_l \otimes \ket{n-\sigma}_{N-l} = \Bigg(\frac{l}{2}-\sigma\Bigg) \ket{\sigma}_l \otimes \ket{n-\sigma}_{N-l}.
\end{equation}

We can now proceed to calculate the generating function 

\begin{equation}\label{eq:gen_fun_calc}
\bra{\Tilde{n}}e^{i \theta \hat{M}(l)}\ket{n} = \frac{1}{\sqrt{\binom{N}{\Tilde{n}}\binom{N}{n}}} \sum_{\sigma,\Tilde{\sigma}} e^{i \theta \Big(\frac{l}{2}-\sigma\Big)} \binom{l}{\sigma}\delta_{\Tilde{\sigma},\sigma}
\binom{N-l}{n-\sigma} \delta_{\Tilde{n}-\Tilde{\sigma},n-\sigma}
\end{equation}

replacing equation \ref{eq:gen_fun_calc} in equation \ref{eq:gen_fun} we get the final expression for generating function of the PDF 

\begin{equation}\label{eq:gf_fin}
    G_l(\theta,t) = \sum_{n = 0}^N \sum_{\sigma = max(0,n-N+l)}^{min(l,n)} \frac{1}{\binom{N}{n}}\abs{X_n(t)}^2  e^{i \theta (\frac{l}{2}-\sigma)} \binom{l}{\sigma}\binom{N-l}{n-\sigma}.
\end{equation}

The PDF can be calculated from this expression of generating function by a simple Fourier transformation as given in equation \ref{eq:fcs9}. 

In figure \ref{fig:fulcon_com} we plot the PDF after quantum quench to two different final transverse field at three different time slices and the evolution of formation probabilities. The results show an excellent match between the exact diagonalization results and TDVP for both quenches.

\section{Long range Ising Hamiltonian as an MPO}\label{app:MPO_ERR}
\begin{figure}[hb]
    \centering 
  \includegraphics[width=0.60\linewidth]{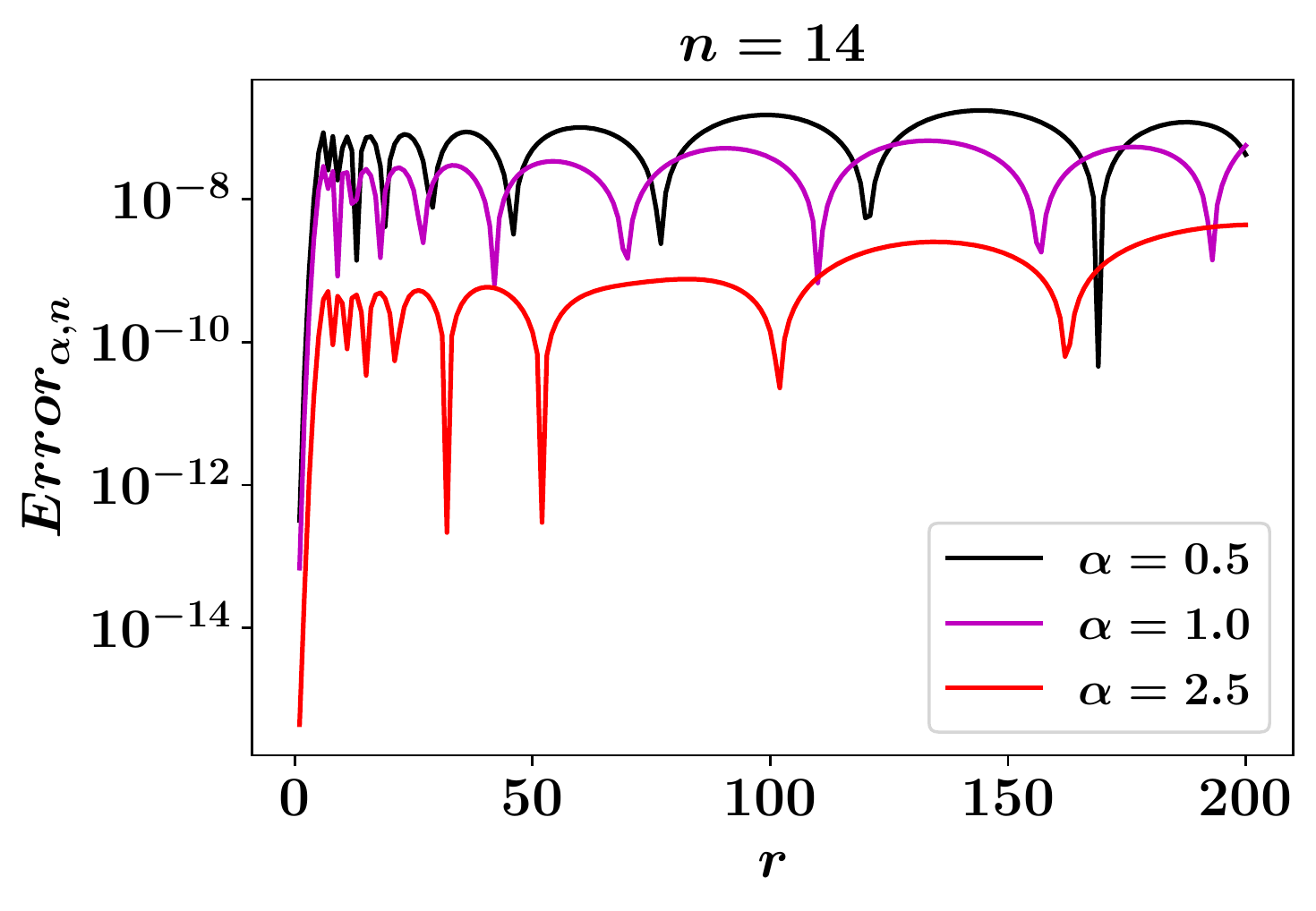}
\caption{Relative Error for representating the power law decay as the sum of exponential for three representative values of interaction range $\alpha = \{0.5,1.0,2.5\}$.}
\label{fig:MPO_error}
\end{figure}

It is straightforward to represent a long range Hamiltonian with exponentially decaying interaction as an MPO \cite{DMRG_MPS}. Representing a long range Hamiltonian with power law decay requires first representing the power law decay as the sum of exponential \cite{MPO1} and then representating the sum of exponentials as an MPO \cite{MPO2}. The goodness of this representation depends on how precisely do we represent the power law decaying function as the sum of exponentials which is quantified by a metric $\textit{Error}_{\alpha,n}$

\begin{equation} \label{eq:MPO}
\textit{Error}_{\alpha,n} = \left|\frac{1}{r^{\alpha}} - \sum_{i=1}^{n} x_i\lambda_i^r\right|
\end{equation}

where the number of exponentials in the sum $n$ determines the precision of fitting. We observe that for system size $L = 200$ the relative error is $10^{-7}$ or smaller~[{\it cf.} Fig.~\ref{fig:MPO_error}] with $n = 14$.  

\section{Convergence with bond dimension}\label{app:convergence}

\begin{figure}
    \centering 
  \includegraphics[width=\linewidth]{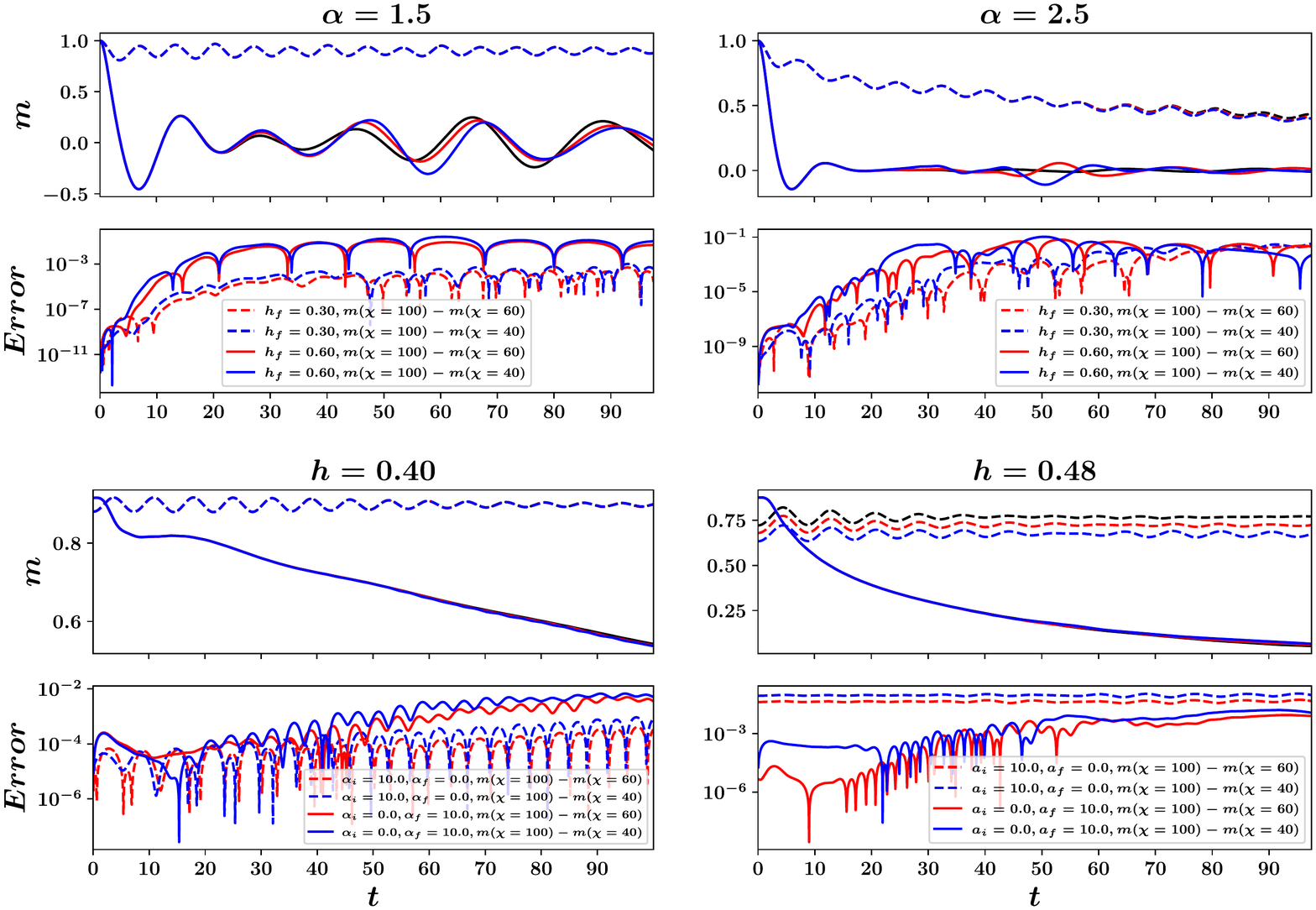}
\caption{Convergence of subsystem magnetization and relative errors with increasing bond dimension, $\chi = 40$ (blue), $\chi = 60$ (red), and $\chi=100$ (black). The top row is for $\alpha = \{1.5,2.5\}$ and $h_f = 0.30$(colored dotted) , and $h_f=0.60$(colored bold). The bottom row is for $h = \{0.40,0.48\}$ and quench from $\alpha = 10.0$ to $\alpha = 0.0$ (colored dotted) $\alpha = 0.0$ to $\alpha = 10.0$ (colored bold)}
\label{fig:chi_error_1}
\end{figure}

To ensure the data generated by the simulations are correct we need to check the convergence of the errors with increasing bond dimension. In TDVP the bond dimension is responsible for projection error~\cite{Time_MPS}, which is a primary source of error. To check that the errors converge with increasing bond dimension we compare the time evolution of subsystem magnetization and relative errors for some representative cases of quantum quenches for $\chi = \{40,60,100\}$ in figure \ref{fig:chi_error_1}. Quenches along the transverse field show qualitatively similar behavior, the relative error converges and becomes flat in a long time limit for all values of post quench parameter. Furthermore, for times up to $25$, which is the maximum time reached for most of the results in the main text, the relative errors are smaller than $O(10^{-3})$. For quenches of the interaction range with $h = 0.40$ we observe a similar behavior. For $h = 0.48$, the quench from $\alpha_i = 10.0$ to $\alpha_f = 0.0$ shows a markedly different behavior. The magnetizations at different $\chi$ maintains a constant shift from one another right from $t = 0$ throughout the evolution. This is because the point $\alpha = 10.0$ and $h = 0.48$ is close to the critical point of the equilibrium phase diagram where we see a logarithmic divergence of the entanglement entropy~\cite{ent_crit} and DMRG generate a considerable relative error while initializing the system in the ground state. This initial error simply gets propagated throughout the time evolution. The relative errors in opposite direction behaves normally attaining a flat region after oscillations. 

\begin{figure}
    \centering 
  \includegraphics[width=\linewidth]{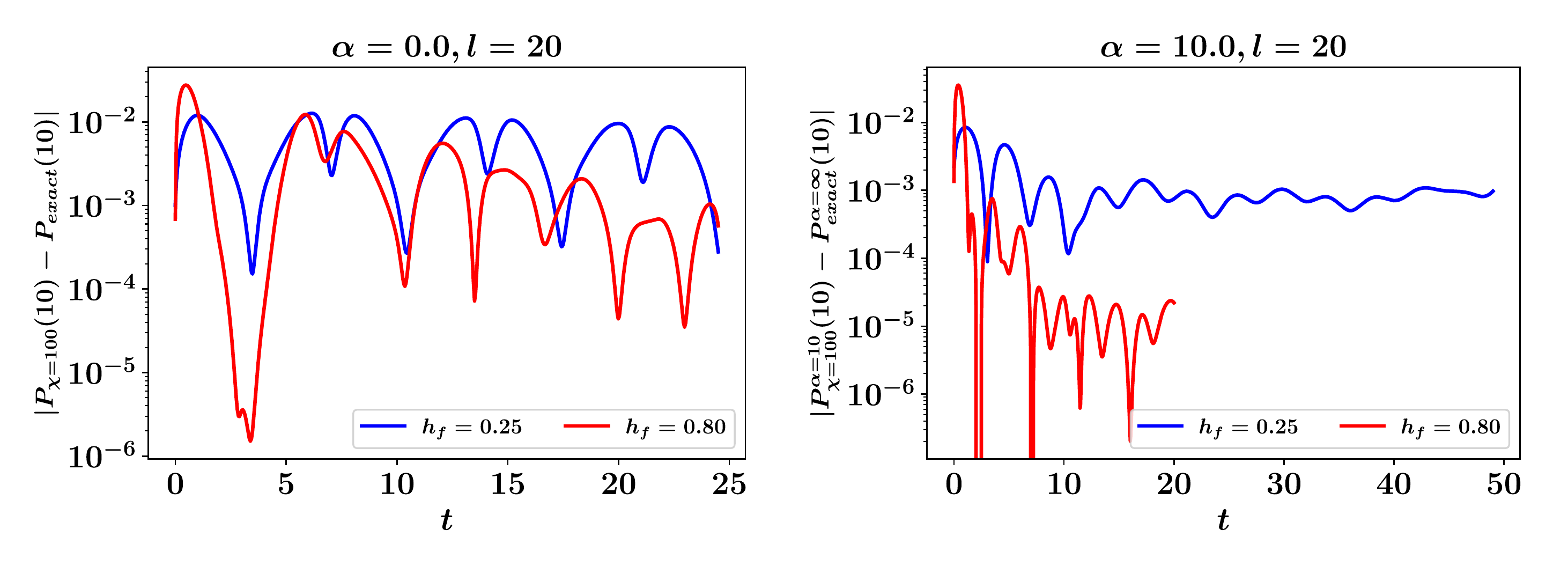}
\caption{Errors in emptiness formation probability computed with bond dimension $\chi = 100$ with respect to the exact analytical and numerical results at two extremes of the long range Ising model for quenches to two values of final transverse fields. On the left panel the error is computed by subtracting the exact diagonalization results with the TDVP results, refer to figure \ref{fig:fulcon_com}. On the right panel the error is computed by subtracting the exact analytical results with the TDVP results computed for sufficiently large interaction range, $\alpha = 10$, refer to figure \ref{fig:short_com}.}
\label{fig:EFP_error}
\end{figure}

In figure \ref{fig:EFP_error} we also plot the error in emptiness formation probability with respect to the exact analytical and numerical results at the two solvable extremes of the model. We observe that these errors are smaller than the order of $O(10^{-2})$ for the overall time evolution.  

\end{appendix}

\clearpage

\bibliography{SciPost_Example_BiBTeX_File.bib}

\nolinenumbers

\end{document}